\documentclass{aa}

\usepackage{txfonts}
\usepackage{graphicx}
\usepackage{amsmath}
\usepackage{color}
\usepackage{gensymb}

\bibpunct{(}{)}{;}{a}{}{,} 

\usepackage{hyperref}

\newcommand{\equ}[1]{Eq.~(\ref{#1})}
\newcommand{\equs}[2]{Eqs.~(\ref{#1})-(\ref{#2})}
\newcommand{\Equ}[1]{Equation~(\ref{#1})}
\newcommand{\Equs}[2]{Equations~(\ref{#1})-(\ref{#2})}
\newcommand{\f}[2]{\frac{#1}{#2}}
\newcommand{\rsun}{R_\odot}
\newcommand{\lat}{\lambda}

\newcommand{\ylm}{Y_{\ell}^m}

\newcommand{\psilm}{\psi_{\ell}^m}
\newcommand{\psilmtilde}{\tilde{\psi}_{\ell}^m}

\newcommand{\Nlm}{N_\ell^m}
\newcommand{\rlm}{r_{\ell}^m}
\newcommand{\slm}{s_{\ell}^m}
\newcommand{\tlm}{t_{\ell}^m}
\newcommand{\Rlm}{\vec{R}_{\ell}^m}
\newcommand{\Slm}{\vec{S}_{\ell}^m}
\newcommand{\Tlm}{\vec{T}_{\ell}^m}

\newcommand{\grad}[1]{\vec{\nabla}#1}

\newcommand{\Rsun}{R_\odot}
\newcommand{\Lsun}{L_\odot}

\begin{document}

\title{Supergranulation and multiscale flows in the solar photosphere}
\subtitle{Global observations vs. a theory of anisotropic turbulent convection}
\author{F. Rincon\inst{1,2}\and T. Roudier\inst{1,2}\and
  A.~A. Schekochihin\inst{3,4} \and M. Rieutord\inst{1,2}}
\institute{Universit\'e de Toulouse; UPS-OMP; IRAP: Toulouse, France \and
           CNRS; IRAP; 14 avenue Edouard Belin, F-31400 Toulouse,
           France \and
The Rudolf Peierls Centre for Theoretical Physics, University of
Oxford, 1 Keble Road, Oxford, OX1 3NP, United Kingdom \and
Merton College, Oxford OX1 4JD, United Kingdom\\  \email{frincon@irap.omp.eu}}

\date{\today}

\abstract{The Sun provides us with the only spatially
  well-resolved astrophysical example of turbulent thermal
  convection. While various aspects of solar photospheric turbulence,
  such as granulation (one-Megameter horizontal scale), are well
  understood, the questions of the physical origin and dynamical
  organization of larger-scale flows, such as the 30-Megameters
  supergranulation and flows deep in the solar convection zone, remain
  largely open in spite of their importance for solar dynamics 
  and magnetism. Here, we present a new critical global observational
  characterization of multiscale photospheric flows and subsequently
  formulate an anisotropic extension of the Bolgiano-Obukhov
  theory of hydrodynamic stratified turbulence that may explain
  several of their distinctive dynamical properties. Our combined
  analysis suggests that photospheric flows in the horizontal range
  of scales between supergranulation and granulation have a typical
  vertical correlation scale of 2.5 to 4 Megameters and operate in a
  strongly anisotropic, self-similar, nonlinear, buoyant dynamical
  regime. While the theory remains speculative at this stage, 
  it lends itself to quantitative comparisons
  with future high-resolution acoustic tomography of subsurface layers
  and advanced numerical models. Such a validation exercise may also
  lead to new insights into the asymptotic dynamical regimes in
  which other, unresolved turbulent anisotropic astrophysical 
  fluid systems supporting waves or instabilities operate.}

\keywords{Sun: photosphere -- Sun: interior --  Convection --
  Turbulence -- Instabilities}
\titlerunning{Observations vs. convection theory of supergranulation
  and multiscale flows in the solar photosphere}

\maketitle

\section{Introduction}
The solar photosphere is the stage of many spectacular
(magneto-)hydrodynamic phenomena and it provides us with a unique
observationally well-resolved example of strongly nonlinear thermal
convection, one of the most common fluid instabilities and transport
processes encountered in nature and astrophysics \citep{kupka07}. While some aspects
of solar thermal turbulence, such as granulation, are now well understood
\citep{nordlund09}, we still lack definitive answers to many important questions,
such as how the turbulence organizes on large scales, and
how it interacts with and amplifies magnetic fields or transports
quantities such as angular momentum or magnetic flux
\citep{miesch05,charbonneau14}. 

The most direct observational characterization of flows connected to
the solar convection zone at scales larger than the granulation has
been achieved through measurements of Doppler-projected velocities at the
photospheric level. In particular, even a simple
visual inspection of Doppler images of the quiet Sun clearly reveals
the pattern of supergranulation flow ``cells'', whose trademark signature 
is a peak around  $\ell\sim 120-130$ (35~Megameters, Mm) 
in the spherical-harmonics energy spectrum of Doppler-projected
velocities \citep{hathaway2000,williams14, hathaway15}. 
The physical origin of this supergranulation is widely debated
\citep{rieutord10a}. In particular, while the idea that
supergranulation-scale flows may just be a manifestation of some form
of thermal convection has a long history, it has often been 
dismissed due to the seeming lack of photometric intensity contrast at the
same scales \citep{langfellner16}. Besides, there is as yet
no general consensus on local helioseismic estimates of the amplitude,
depth and structure of subsurface flows on this scale
\citep{nordlund09,rieutord10a,gizon10,svanda13,degrave14,svanda15} --
or in fact on any scale larger than that \citep[see,
e.g.,][]{hanasoge12,gizon12,hanasoge15,greer15,toomre15,greer16}.
Numerical simulations of the problem have, until recently, also been 
quite limited. While their results have not led to clear-cut
results and conclusions \citep[see reviews
by][]{miesch05,nordlund09,rieutord10a}, many of them
suggest that meso- to supergranulation-scale flows 
have a convective (buoyant) origin
\citep[e.g.,][]{rincon05,lord14,cossette16}. 

In this article, we attempt to make further progress on this problem
using the high-quality observations of the solar photosphere provided
by the Solar Dynamics Observatory (SDO) and a new theoretical
analysis. In Sect.~\ref{obs},  we present a global 
observational analysis of multiscale
photospheric vector flows reconstructed from quasi-full disc Doppler
and photometric data from SDO using a tracking technique. This
analysis subsequently leads us in Sect.~\ref{theory} to formulate a
theory of anisotropic turbulent convection that may explain several
distinctive dynamical properties of these flows. As further argued 
in Sect.~\ref{consistency}, the combined results consistently suggest
that photospheric flows in the horizontal range of scales in between 
supergranulation and granulation operate in a strongly anisotropic,
self-similar, buoyant dynamical regime. Connections between these
results and earlier work, as well as future perspectives, are discussed
in Sect.~\ref{discussion}. The main text of the article is
focused on results. The technical details of the data
processing and analysis are provided in the two Appendices.

\section{Observations\label{obs}}
Our observational analysis is based on 24 hours of uninterrupted
high-resolution white-light intensity and Doppler observations of the
entire solar disc by the HMI instrument aboard the SDO satellite
\citep{scherrer12,shou12}. The data was obtained on October 8, 2010
starting at 14:00:00 UTC, one of the quietest periods of solar activity since
the launch of SDO. Our analysis is distinct and independent of the
recent studies by \cite{williams14,langfellner15,hathaway15}, and
both confirms and extends some of their results. 

\subsection{Data analysis}
\subsubsection{Velocity field reconstruction}
\begin{figure}
\centering
\includegraphics[width=\columnwidth]{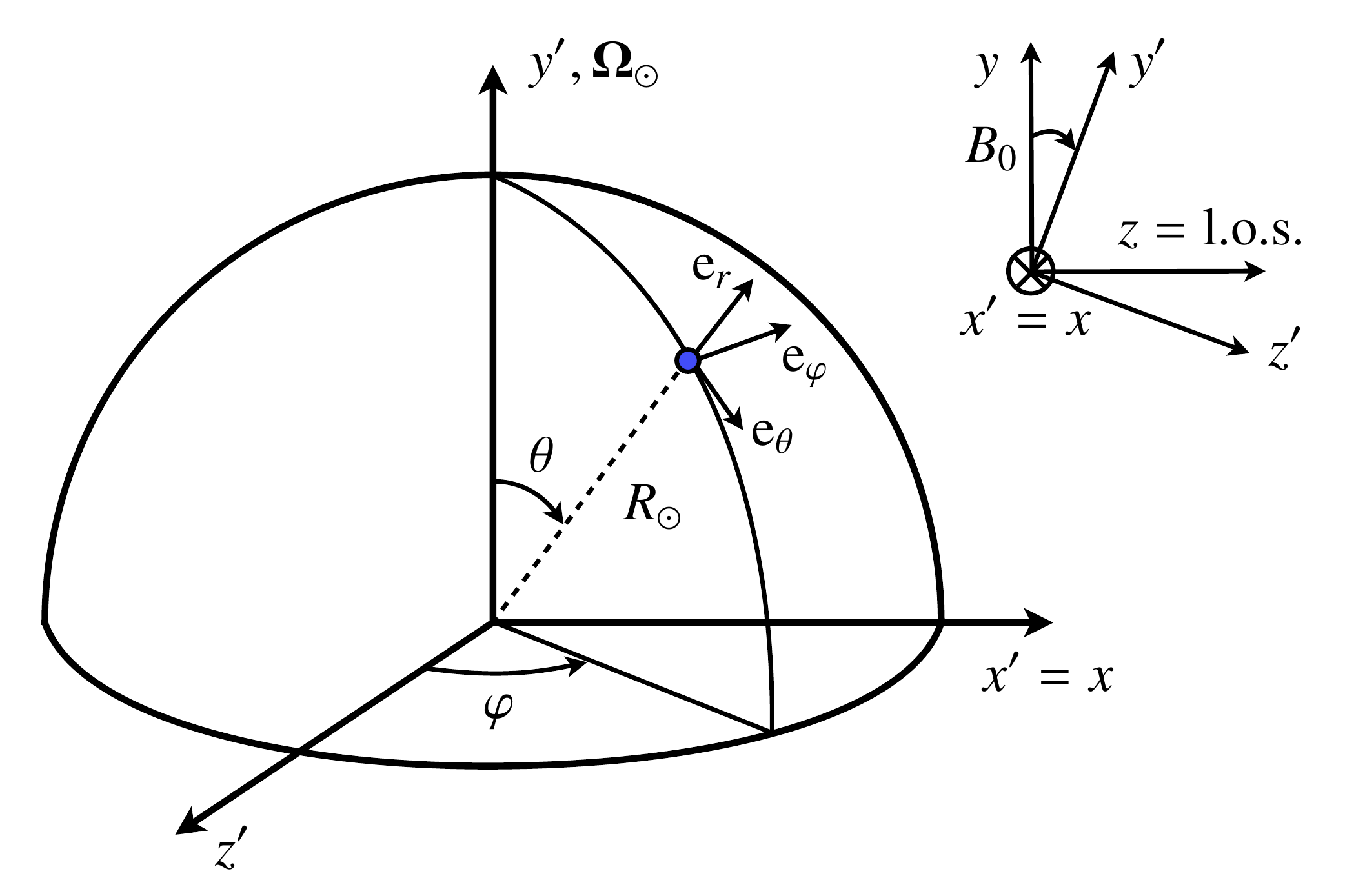}
\caption{\label{figcoord}Coordinate systems used
  in the reconstruction of the 3D photospheric velocity field.}
\end{figure}

Our first objective is to reconstruct the
three components $(u_r,u_\theta,u_\varphi)$ of the photospheric vector
velocity field in the angular degree range $20<\ell<850$ from
available observations. To this end, we perform a
Coherent Structure Tracking analysis \citep[CST,][]{roudier12} of
photometric structures such as granules. This technique provides us
with the projection of the photospheric velocity field $(u_x,u_y)$ onto
the plane of the sky/CCD matrix (Fig.~\ref{figcoord}). We can then 
combine the results with Doppler data, which provides the out-of-plane
component $u_z$ of the velocity field, to calculate the full vector
velocity field at the surface in solar spherical coordinates,
\begin{eqnarray}
u_r(\theta,\varphi) & =& \sin\theta \sin\varphi\,  u_{x} \nonumber\\
&& +\,(\cos \theta \cos B_0 - \sin \theta \cos \varphi \sin B_0)\, u_{y}\nonumber\\
&& +\,(\sin \theta  \cos \varphi \cos B_0 + \cos \theta  \sin B_0)\,
u_{z}~,\label{ur}\\
u_\theta(\theta,\varphi) &=& \cos\theta  \sin\varphi \, u_{x} \nonumber\\
&& -\,(\cos \theta  \cos \varphi  \sin B_0 + \sin \theta  \cos B_0 )\,  u_{y} \nonumber\\
&& -\,(\sin \theta  \sin B_0 - \cos \theta  \cos \varphi  \cos B_0 ) \, u_{z}~,\label{utheta} \\
u_\varphi(\theta,\varphi) &=& \cos \varphi\, u_{x} \nonumber\\
&& +\,\sin \varphi \sin B_0 \, u_{y} \nonumber\\
&& -\,\sin \varphi \cos B_0 \, u_{z}~,\label{uphi}
\end{eqnarray}
where $B_0=6.3\degree$ is the heliographic latitude of the central
point of the solar disc at the time of observation. The reduction and
filtering of the raw photometric and Doppler data required to
build a consistent data set for $(u_x,u_y,u_z)$ is described in
detail in App.~\ref{reconstruct}.

\subsubsection{Spherical-harmonics analysis}
The reconstructed vector field $\vec{u}(\theta,\varphi)=(u_r,u_\theta,u_\varphi)$ 
is expanded in terms of vector spherical harmonics as
\begin{equation}
\label{vectorexpansion}
  \vec{u}(\theta,\varphi)=\sum_\ell\sum_{m=-\ell}^\ell \rlm\,
\Rlm(\theta,\varphi)+\slm\,\Slm(\theta,\varphi)+\tlm\,\Tlm(\theta,\varphi)~,
\end{equation}
where $\Rlm=  \ylm \vec{e}_r$, $\Slm=\Rsun\grad{\ylm}$, and
$\Tlm=-\vec{\Rsun}\times\grad{\ylm}$. The  spectral coefficients $\rlm$
describe the radial flow component, and the coefficients  $\slm$ and
$\tlm$ describe the spheroidal (divergent) and toroidal
(vortical) parts of the horizontal flow components, respectively. 
Global maps of the horizontal divergence and radial vorticity at the
surface are obtained by mapping back $-\ell(\ell+1)\slm/R_\sun$
and $\ell(\ell+1)\tlm/R_\sun$ to physical space. 
The corresponding one-dimensional radial, spheroidal and toroidal
energy spectra are given by
\begin{eqnarray}
\label{Er}
E_r(\ell) & = & \sum_{m=-\ell}^\ell |\rlm|^2~,\\
\label{Es}
E_S(\ell) & = & \sum_{m=-\ell}^\ell \ell\,(\ell+1)\,|\slm|^2~,\\
\label{Et}
E_T(\ell) & = & \sum_{m=-\ell}^\ell \ell\,(\ell+1)\,|\tlm|^2~,
\end{eqnarray}
respectively. The $\ell(\ell+1)$ prefactors in $E_S$ and $E_T$ result
from the definition of the spheroidal and toroidal vector spherical
harmonics basis vectors. 

The technical aspects of the harmonic-transforms procedure of SDO/HMI data
are described in detail in App.~\ref{harmonicanalysis}. Here, it suffices
to mention that the data must first be apodized because we only
see one side of the Sun, and we must eliminate near-limb regions where the CST
analysis is not reliable. Besides, as will be shown below, the radial
component of the flow is very weak and can only be tentatively
determined using data close to the disc center to limit contamination 
by the intrinsic algorithmic noise of the CST in the deprojection
process. We therefore use an apodizing
window with a $15^{\degree}$ opening heliocentric angle from the disc center
to compute the spectra of the radial and line-of-sight (Doppler)
velocity fields, and a window with a $60^{\degree}$ opening angle to
compute the spectra of the horizontal velocity field. The general
mathematical definition of these windows is provided in \equ{eq:rhomask}.

\begin{figure*}
\vbox{\hspace{1cm}a)

\centering
 \includegraphics[width=\textwidth]{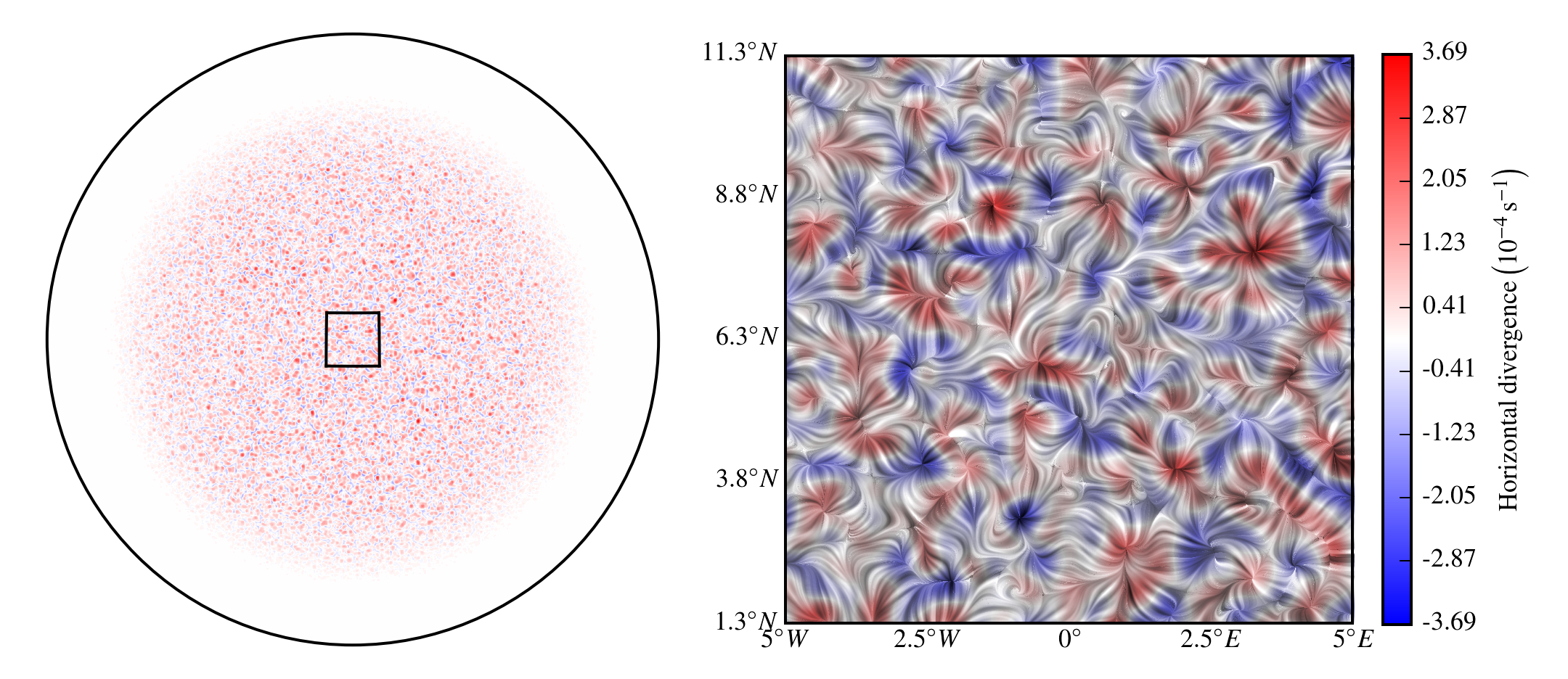}
}
\vbox{\hspace{1cm} b)

\centering
\includegraphics[width=\textwidth]{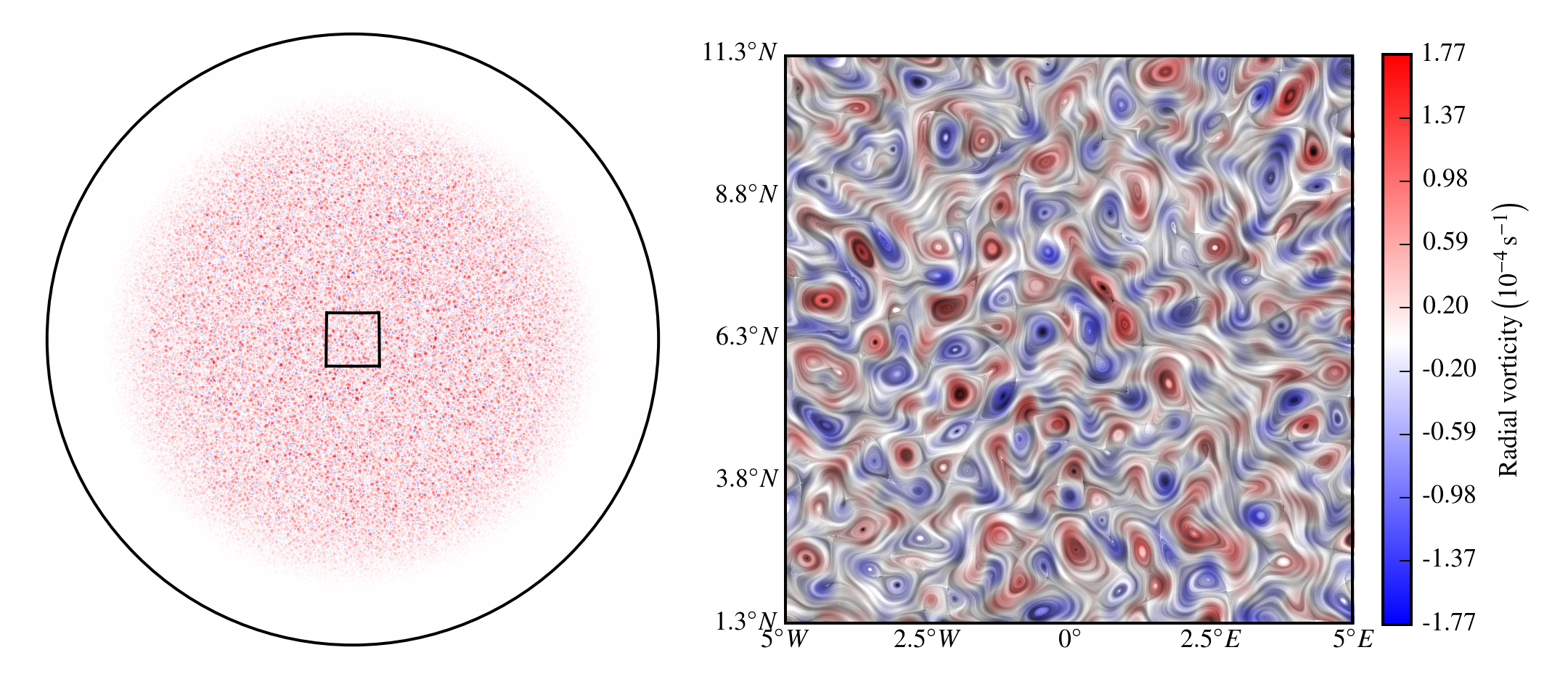}
}
\caption{\label{snapshotsdiv}a) Left: quasi-full-disc map of the horizontal
  divergence $-\ell(\ell+1)\slm/R_\sun$ (color contours) of the
  30-minute-averaged horizontal surface velocity field at 14:00:00
  UTC, apodized beyond $60^{\degree}$ from the disc center.
  Right: local map at the disc center
  of the divergence field convolved by 
  horizontal-flow Lagrangian trajectories using a line-integral
  convolution visualization technique. The map in the right panel
  is a zoom on the square patch in the left panel. b)
  Left: quasi-full-disc map of the vertical
  vorticity $\ell(\ell+1)\tlm/R_\sun$ (color contours) of the
  30-minute-averaged horizontal surface velocity field at 14:00:00
  UTC, apodized beyond $60^{\degree}$ from the disc center.
  Right: local map at the disc center of the same vorticity field
  convolved by Lagrangian trajectories computed from the vortical part
  of the horizontal flow only (the same line-integral convolution visualization
  technique is used). The map in the right panel is a zoom on
  the square patch in the left panel.}
\end{figure*}

\subsection{Results}
\subsubsection{Velocity field maps}
Figure~\ref{snapshotsdiv} shows maps of the horizontal divergence
field $\vec{\nabla}_h\cdot\vec{u}_h$ and vertical vorticity field
$\vec{e}_r\cdot\left(\vec{\nabla}_h\times\vec{u}_h\right)$ derived from a single
snapshot of the 30-minute-averaged horizontal surface velocity field
$(u_\theta,u_\varphi)$ using the spherical-harmonics spheroidal-toroidal
decomposition into divergent and vortical motions. The pattern is
dominated by supergranulation, whose divergent morphology is revealed
in the close-up panel using a line-integral convolution technique
highlighting horizontal-flow Lagrangian trajectories
\citep{cabral93}. The vortical component of the flow is significantly
weaker than the divergent component, as indicated by the associated rate
of strains reported in the color bars (and further demonstrated in the
spectral analysis presented below). Figure~\ref{urdiv} shows a 
local map in the same disc-center zone of the 24h-averaged $u_r$,
superimposed with horizontal flows averaged over the same time scale. 
There is a noticeable correlation between the weak upflows (downflows)
and horizontal divergences (convergences) at the supergranulation scale.

\begin{figure}
\includegraphics[width=\columnwidth]{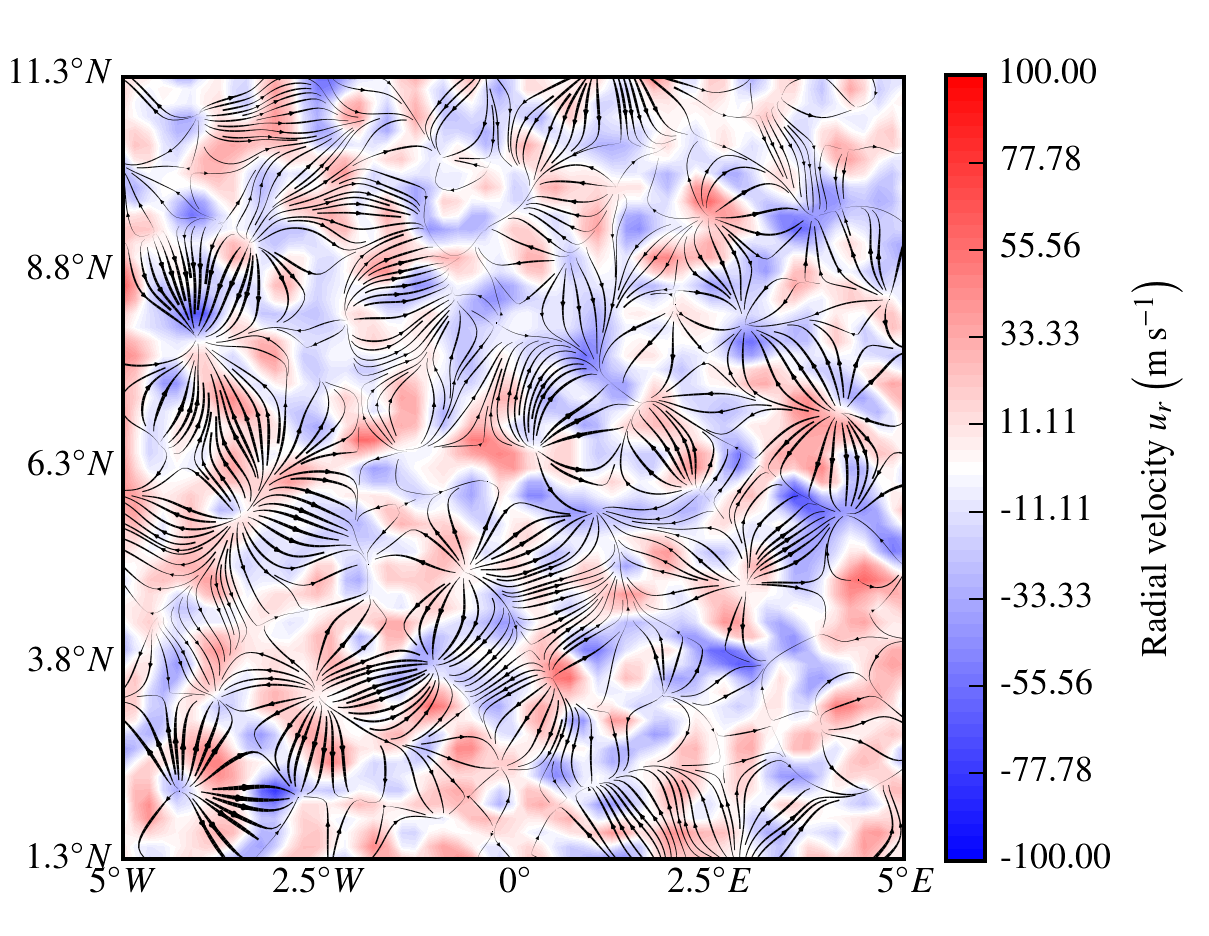}
\caption{\label{urdiv}Local map at the disc center of the 24h-averaged radial
  (color contours) and horizontal spheroidal flow components (arrow field).}
\end{figure}

\subsubsection{Energy spectra}
\label{spectra}
\begin{figure}
\centering
 \includegraphics[width=\columnwidth]{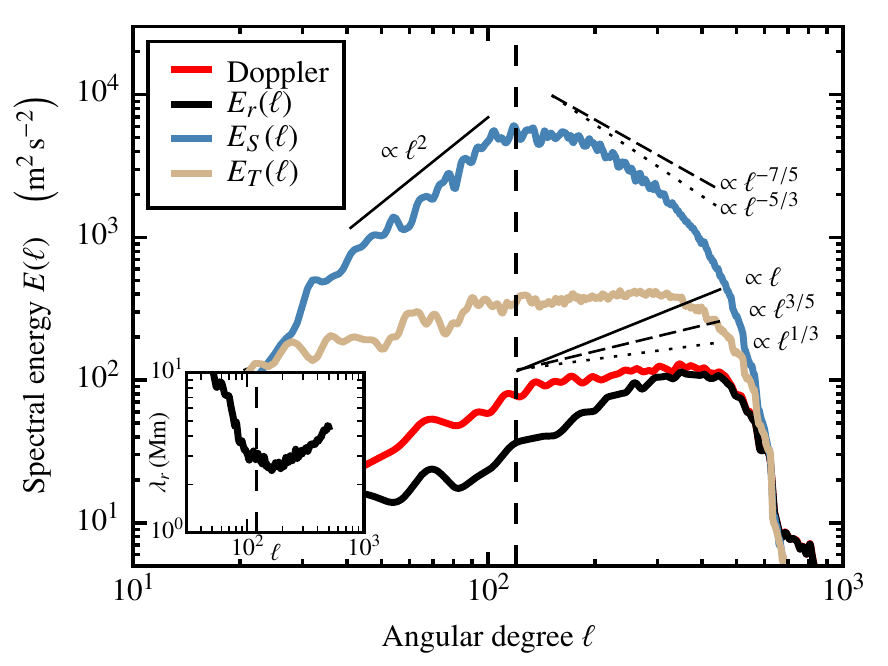}
\caption{\label{flowspectra}
Spherical-harmonics energy spectra (see \equs{Er}{Et}) of maps of the
30-minute-averaged (top) velocity field derived from the joint
CST-Doppler analysis. The thin lines indicate different possible
spectral slopes (see Sect.~\ref{theory}).  Vertical dashed line:
$\ell=120$. Inset: vertical scale estimate.}
\end{figure}

Figure~\ref{flowspectra} shows the spherical-harmonics radial,
spheroidal and toroidal energy spectra $E_r(\ell)$, $E_S(\ell)$, and
$E_T(\ell)$ of the 30-minute-averaged velocity-field components.
The horizontal divergent component is energetically
dominant at all scales probed: the corresponding spheroidal spectrum
$E_S(\ell)$ exhibits a clear peak at the supergranulation scale 
$\ell\sim 120$, corresponding to a typical r.m.s. flow velocity of
$400~\textrm{m}\,\textrm{s}^{-1}$. The r.m.s. radial flow is
$20-30~\textrm{m}\,\textrm{s}^{-1}$ at $\ell\sim 120$.
The spectra of the different flow components behave differently
for $\ell>120$: the spheroidal $E_S(\ell)$ and radial
$E_r(\ell)$ spectra respectively decay and increase with decreasing
scale, while the toroidal spectrum $E_T(\ell)$ is almost flat. The
exact shapes of the weaker and noisier $E_r(\ell)$ and $E_T(\ell)$ are
moderately sensitive to the processing and averaging of the data,
but the general trend described here is robust (as explained in
App.~\ref{ellnufiltering}, it is very difficult to separate the
  Doppler contributions of the large-scale spectral tail of granulation from
  those of the radial component of the large-scale motions otherwise
detected by the CST. Consequently, and while we did our best to remove 
the large-scale contribution of granulation from the Doppler data, our
results for $E_r(\ell)$ in the range $120<\ell<500$ still probably slightly
overestimate the actual amplitude of the radial component of flows not
associated with granulation. We also find that $E_r(\ell)$ lies
somewhere between $\ell^{1/2}$ and $\ell$ power laws, depending on the
exact filtering procedure applied to the Doppler signal). The
spectral falloff at $\ell> 500$ is due to
the intrinsic 2.5--5~Mm CST spatial resolution down to which the
vector-field reconstruction was performed.  The different power laws
that these apparently self-similar spectra follow for $\ell<500$ will
be discussed in Sects.~\ref{scalinglaws} and \ref{discussscalinglaws}.

\subsubsection{Vertical correlation scale\label{obsvestimate}}
Finally, we estimate the typical vertical correlation scale $\lambda_r$
of these subsonic flows as a function of horizontal scale
$\lambda_h=2\pi/k_h$ (or of the angular degree $\ell=k_h\Rsun$). This
can be obtained from the mass conservation relation 
\begin{equation}
\label{massconservation}
\frac{u_h}{\lambda_h}\sim \frac{u_r}{\lambda_r}~,
\end{equation}
where $u_h$ and $u_r$ are the horizontal and vertical velocity fluctuations
(because the medium is stratified, here $\lambda_r$ serves as an
\textit{a priori} proxy notation for either the radial wavelength of
the perturbations or the local density scale height, whichever
one is smaller). We therefore define 
\begin{equation}
\lambda_r(\ell)=\frac{2\pi\Rsun}{\ell}\sqrt{\frac{E_r(\ell)}{E_S(\ell)}},
\end{equation}
and plot this quantity in Fig.~\ref{flowspectra} (inset) for
$30<\ell<500$. Remarkably, the typical vertical correlation scale does
not vary significantly in the range $120<\ell<500$
($\left[9,35\right]$~Mm), 
that is $\lambda_r\sim \left[2.5,4\right]$~Mm. Considering the
uncertainties  in the determination of $E_r$ discussed in
Sect.~\ref{spectra},  this scale estimate should probably be
considered as an upper bound rather than as an exact value. Overall,
this scale appears to be comparable with the density scale height
$H_\rho\simeq[1,2]~\mathrm{Mm}$ below the surface, but is somewhat
deeper than the granulation thermal boundary layer \citep{nordlund09}.

\subsection{Main conclusions}
The observational analysis presented above leads us to the following
conclusions. 

\begin{itemize}
\item[(i)] The flow in the range $120<\ell<500$ 
is characterized by divergent (convergent) horizontal
flows correlated with upflows (downflows). This predominantly
cell-like  morphology of the flow suggests that the dynamics
in that range of scales is dominated by buoyancy forces.
\item[(ii)] The ratio of $u_h$ to $u_r$ suggests that
motions are strongly anisotropic $(\lambda_r\ll \lambda_h)$ and
strongly feel the effects of stratification, with the vertical
correlation scale of order the density scale height.
\item[(iii)] The significant energy content of horizontal motions and their
spectral break at the supergranulation scale, followed by the apparent
power-law decay of their spectrum at smaller horizontal scales,
suggest that supergranulation corresponds to the largest
buoyancy-driven scale of the system, below which some form of
nonlinear cascade takes place.
\end{itemize}

\section{Theoretical considerations\label{theory}}
Let us now attempt a theory of anisotropic turbulent convection
that could explain these observations. As argued by
\cite{rincon07}, a good starting point is the Bolgiano-Obukhov
\citep[BO59, see][]{oboukhov59,bolgiano62} phenomenology, which is
based on the following assumptions: (i) a constant spectral flux of buoyancy
fluctuations follows from the temperature/energy equation,
(ii) a balance between the inertial and buoyancy terms
in the momentum equation, and (iii) isotropy of motions.
BO59 predicts a $k^{-7/5}$ thermal spectrum, and a steep
$k^{-11/5}$ velocity spectrum. For the Sun, the latter is ruled 
out by Fig.~\ref{flowspectra} for $120<\ell<500$. This is
not very surprising because, while the first two BO59 assumptions
make sense in the present context, the isotropy assumption must
manifestly be false because $\lambda_r\ll\lambda_h$ (see
Sect.~\ref{obsvestimate}).
 
More generally, we point out that the isotropic
  BO59 scaling laws are not observed either in simulations of
  Rayleigh-B\'enard convection with order-unity vertical to
  horizontal aspect ratio \citep{rincon06,rincon07,kumar14}.
While the conclusions of these previous studies apply to the regime
$kH>1$ (where $k$ is an isotropic wave number), they do not appear to
be directly applicable
to the strongly anisotropic regime $k_hH\ll 1$ characteristic of the
astrophysical problem investigated in the present paper. In any case,
in what follows, we only use the phenomenology of the
original isotropic BO59 theory as a physical motivation to derive distinct
anisotropic scaling laws for the largely unexplored ``long horizontal
wavelength'' regime $k_hH\ll 1$.

\subsection{Theoretical framework}
An anisotropic generalization of BO59 may be derived from the
following set of dynamical equations valid in the Boussinesq 
and anelastic limits:
\begin{eqnarray}
\label{NS}
\frac{\partial\vec{u}}{\partial t}
+\vec{u}\cdot\vec{\nabla}\vec{u} & = & -\vec{\nabla}p +
\frac{\theta}{\Theta_0}g\,\vec{e}_r +\nu\Delta\vec{u}~,\\
\label{thermal}
\frac{\partial\vec{\theta}}{\partial t} +
\vec{u}\cdot\vec{\nabla}\theta &= & -u_r\cdot\nabla_r\,\Theta_0
+\kappa\Delta \theta~, \\ 
\label{div}
\vec{\nabla}\cdot{(\rho_0\vec{u})} & = & 0~.
\end{eqnarray}
Here $\rho_0$ is the mean density, $\vec{u}$ denotes velocity
perturbations,  $p$ denotes pressure fluctuations divided by
$\rho_0$,  $\Theta=\Theta_0+\theta$ is the actual
temperature in the Boussinesq limit, and the potential
temperature $\left(P\rho^{-\gamma}\right)^{1/\gamma}$ in the
anelastic limit, $\Theta_0(r)$ is its horizontally averaged part
and $\theta$ its fluctuations around this average,  $\nu$ is the
kinematic viscosity, $\kappa$ is the thermal diffusivity
($\kappa\gg\nu$ in the Sun), and $\vec{g}=-g\,\vec{e}_r$ is
the acceleration of gravity. 

\subsection{Phenomenology of anisotropic turbulent convection}
The solar convection zone underneath the
thin thermal boundary layer at the surface is in the strongly nonlinear
convection regime, so its thermodynamic profile is very close to
adiabatic and $\Theta_0$ is almost independent of $r$. We may then
argue that the thermal injection term on the
r.h.s. of \equ{thermal} is small compared to the nonlinear
advection term on the l.h.s.. Similarly to the Kolmogorov phenomenology in hydrodynamics
\citep[K41, see, e.g.,][]{davidson}, which postulates a constant flux of
kinetic energy in the inertial range ($\delta u^3/\lambda\sim
\mathrm{const})$, the dynamics in this ``thermal-inertial'' regime is
characterized by a constant flux of $\theta$ variance from scale to scale:
\begin{equation}
\label{SFtheta}
\frac{\delta u_h\,\delta \theta^2}{\lambda_h}\sim\varepsilon_\theta~ = \mathrm{const},
\end{equation}
where $\varepsilon_\theta=\left<\kappa|\nabla\delta
\theta|^2\right>$ is the (time) average dissipation rate of thermal
fluctuations (equal to their injection rate) and $\delta$ denotes
increments of fluctuations between two points separated by a
horizontal separation vector $\vec{\lambda}_h$ at a constant 
radial coordinate. \Equ{SFtheta} effectively corresponds to BO59
assumption (i). This can, in fact, also be formalized on the basis of
Yaglom's exact law for dissipative scalars \citep{monin71,rincon06}, 
which \equ{thermal} obviously satisfies if $\nabla_r\Theta_0=0$. 
Similarly to BO59 assumption (ii), we seek a
buoyancy-dominated  dynamical regime balancing the inertial and
buoyancy terms in \equ{NS}, but we drop the isotropy
assumption, expecting instead that $\lambda_r\ll \lambda_h$. The
result of this is
\begin{eqnarray}
\label{SFz}
\frac{\delta u_h\, \delta u_r}{\lambda_h} & \sim & \left(\frac{\lambda_r}{\lambda_h}\right)^2
g\frac{\delta\theta}{\Theta_0}~.
\end{eqnarray}
The small $(\lambda_r/\lambda_h)^2$ prefactor on the r.h.s. accounts
for the partial cancellation of buoyancy by the pressure gradient in
the vertical direction, as a result of the horizontal redistribution 
of buoyancy work by pressure forces (this prefactor also appears in the
dispersion relation of gravity waves in the limit $\lambda_r\ll
\lambda_h$). 

\subsection{Scaling laws}
\label{scalinglaws}
\Equs{SFtheta}{SFz}, combined with the mass conservation relation
\equ{massconservation}, lead to the following self-similar scaling relations,
\begin{eqnarray}
\delta u_r & \sim  & \left(\frac{\varepsilon_\theta}{\Theta_0^2}\right)^{1/5}g^{2/5}
  \lambda_r^{7/5}\lambda_h^{-4/5}~,\label{deltauz}\\
\delta u_h &\sim & \left(\frac{\varepsilon_\theta}{\Theta_0^2}\right)^{1/5}g^{2/5}\lambda_r^{2/5}\lambda_h^{1/5}~,\label{deltauh}\\
  \delta \theta/\Theta_0 & \sim
  &\left(\frac{\varepsilon_\theta}{\Theta_0^2}\right)^{2/5}g^{-1/5}\lambda_r^{-1/5}\lambda_h^{2/5}~.\label{deltaT}
\end{eqnarray}
As a consistency check, we can see that the isotropic BO59
prediction $\delta u \sim \lambda^{3/5}$, $\delta \theta \sim
\lambda^{1/5}$ is recovered if we set $\lambda_r=\lambda_h$.
Here however, we are instead interested in a strongly stratified
regime in which the typical vertical scale height $H_\rho$ of the
system is much smaller than $\lambda_h$. We therefore treat
$\lambda_r$ as a simple parameter equal to $H_\rho$ in
\equs{deltauz}{deltaT}, as also suggested by our observational
result of Sect.~\ref{obsvestimate}. Doing this results in the
following power-law energy spectra as functions of horizontal wave
number $k_h=\ell/\Rsun$,
\begin{eqnarray}
\label{vscaling}
E_r(k_h) & \sim &
\left(\frac{\varepsilon_\theta}{\Theta_0^2}\right)^{2/5}g^{4/5}H_\rho^{14/5}k_h^{3/5}~,\\
\label{hscaling}
E_h(k_h) & \sim &
\left(\frac{\varepsilon_\theta}{\Theta_0^2}\right)^{2/5}g^{4/5}H_\rho^{4/5}k_h^{-7/5}~,\\
\label{tscaling}
E_\theta(k_h) & \sim & \left(\frac{\varepsilon_\theta}{\Theta_0^2}\right)^{4/5}g^{2/5}H_\rho^{-2/5}k_h^{-9/5}~.
\end{eqnarray}
We note that in the regime $k_hH_\rho\ll 1$, relevant to the astrophysics
  problem investigated here, these scalings are distinct
  from the classical isotropic BO59 prediction.

\section{Comparison of theory with observations\label{consistency}}

\subsection{Self-similar buoyant dynamics}
Physically, the theory proposed above describes a situation in which
buoyant thermal fluctuations stochastically injected through the
surface boundary layer drive nonlinear anisotropic convective motions
(plumes) in the adiabatic layer. By mixing these thermal fluctuations,
turbulent velocity fluctuations drive a horizontal cascade of buoyancy
down to thermal dissipative scales, resulting in statistically self-similar
buoyant dynamics. This picture appears to be very much in line with
the observed phenomenology of flows at scales larger than the
granulation, notably the hierarchical, bubbling recurrent dynamics of
``trees and families of fragmenting granules'' documented by
\cite{roudier03,roudier09,roudier16}. The dynamics are 'quasi-2D' in
the sense that there is no refinement of the vertical scale as the
horizontal cascade proceeds, but it is 3D in the sense that finite
$\lambda_r \sim H_\rho$ is required to ensure mass conservation.

\subsection{Spectral amplitudes vs. theoretical estimates\label{thestimate}}
We now test for consistency between the observed spectral
amplitudes and our theoretical scaling estimates derived
by assuming that the observed  fluctuations have a convective
origin. A fit of the observed $E_S(\ell)\sim E_h(k_h)/\Rsun$ to
\equ{hscaling} in the range $120<\ell<500$ for $H_\rho\simeq 2~\mathrm{Mm}$ gives
\begin{equation}
\label{epsobs}
\left(\frac{\varepsilon_\theta}{\Theta_0^2}\right)_\mathrm{obs}\simeq 10^{-9}\,\mathrm{s}^{-1}~.
\end{equation}
On the other hand, a theoretical dimensional
estimate of the same quantity can be obtained directly in terms of 
the solar luminosity $\Lsun$ and internal structure of the Sun.
The convective flux as a function of horizontal scale,
as estimated from \equs{deltauz}{deltaT}, is
\begin{equation}
\label{convectiveflux}
F_c\sim \rho_0\,c_v\,\delta\theta\,\delta u_r \sim \rho_0\,c_v\,\Theta_0\left(\frac{\varepsilon_\theta}{\Theta_0^2}\right)^{3/5}g^{1/5}\lambda_r^{6/5}\lambda_h^{-2/5}~,
\end{equation}
where $c_v$ is the specific heat at constant volume. We see that this
flux increases with decreasing horizontal scale (the
  horizontal kinetic energy, on the other hand,
  increases with increasing horizontal scale in this regime. This is a
  consequence of the different anisotropic scalings for $u_r$ and
  $u_h$, and suggests that supergranulation-scale convection, while 
dominant in terms of horizontal kinetic energy, does not transport a
significant fraction of the solar flux). However, this scaling
theory is only applicable for $\lambda_h\gg\lambda_r$.  As
$\lambda_h$ becomes of the order of $\lambda_r\sim H_\rho$, there
should be a crossover to either the standard isotropic Bolgiano regime
$\delta u\sim \lambda^{3/5}$, $\delta \theta\sim\lambda^{1/5}$, or
directly to the Kolmogorov regime $\delta u\sim \lambda^{1/3}$,
$\delta \theta\sim\lambda^{1/3}$ \citep{kumar14}. In both cases,
the cascade proceeds in both horizontal and vertical directions, with
$F_c\sim \lambda^{4/5}$ for BO59 or $F_c\sim \lambda^{2/3}$ for K41,
that is, the convective flux decreases with decreasing scale $\lambda$. 
We therefore argue that the convective flux peaks at the
isotropization scale $\lambda_h\sim\lambda_r\sim H_\rho$, and 
require that $F_c$ at this scale be of the order of the total solar
flux,
\begin{equation}
\label{flux} 
\rho_0\,c_v\,\Theta_0\,\left(\frac{\varepsilon_\theta}{\Theta_0^2}\right)^{3/5}g^{1/5}H_\rho^{4/5}\sim
\frac{\Lsun}{4\pi\rsun^2}~,
\end{equation}
where $\Lsun\sim 3.8\times 10^{26}~\mathrm{W}$ is the solar
luminosity. 
 Using $\rho_0\simeq 10^{-2}\,\mathrm{kg}\,\mathrm{m}^{-3}$,
$\Theta_0\simeq 1.9 \times 10^4\,\mathrm{K}$, $c_v\simeq 10^5
\mathrm{J}\,\mathrm{kg}^{-1}\mathrm{K}^{-1}$ \citep{stixbook}
 and $H_\rho\simeq 2\,\mathrm{Mm}$ in \equ{flux},
we obtain the theoretical estimate
\begin{eqnarray}
\left(\frac{\varepsilon_\theta}{\Theta_0^2}\right)_\mathrm{th} & \sim &
\left(\frac{\Lsun}{4\pi\rsun^2\,\rho_0\,c_v\,\Theta_0\,g^{1/5}H_\rho^{4/5}}\right)^{5/3}\nonumber\\
& \sim & 4\times 10^{-9}\,\mathrm{s}^{-1}~,
\end{eqnarray}
which, given the approximate nature of the calculation and the neglect
of geometric prefactors in \equs{deltauz}{deltaT} and
\equs{vscaling}{tscaling}, is relatively close
 to our observational estimate in \equ{epsobs}
(this cascade time scale also appears to be similar to the
typical plume  thermal diffusion time scale 
in the Sun \citep{miesch05,cossette16}, that is, the ``turbulent''
diffusivity $H_\rho \delta u (H_\rho)$ at scale $H_\rho$ is of the
order of the molecular diffusivity $\kappa$). Overall, the observed
 amplitudes therefore seem consistent with a convective origin of
 the turbulence. We note, finally, that if this estimate holds, then
 \equ{deltaT} suggests relative temperature fluctuations $\delta
 \theta/\Theta_0$ of approximately $1\%$ at supergranulation 
scales in the first few Mm below the surface, consistent with
helioseismic inferences \citep[e.g.,][]{duvall97}.
  
\subsection{Spectral power laws}
\label{discussscalinglaws}
The horizontal spectral scaling laws predicted by
our theory for the vertical and  horizontal velocities, 
 $E_r(\ell)\sim E_r(k_h)/\rsun\propto \ell^{3/5}$ (\equ{vscaling}) and
   $E_S(\ell)\sim E_h(k_h)/\rsun\propto \ell^{-7/5}$ (\equ{hscaling}), 
are broadly consistent with our observations, as shown in
Fig.~\ref{flowspectra} by the power laws in the relevant range of
scales. Different tentative scalings $E_h(\ell) \propto \ell^{-5/3}$
and $E_r(\ell)\propto \ell^{1/3}$, derived by naively postulating K41
scalings for the horizontal velocity field and using the mass conservation
\equ{massconservation} with $\lambda_r\sim H_\rho$ to obtain the scaling of the
vertical velocity component, are shown for comparison (dotted
lines). Such scalings cannot be ruled out given the closeness of their
spectral exponents to those of our theory
and the aforementioned sensitivity in the determination of
$E_r(\ell)$ to the details of data processing. However, we see no
reasonable physical argument as to why or how  such an ``anisotropic
K41'' regime would actually be realized in this environment, in
contrast with the nonlinear buoyancy regime described
above which naturally produces divergent horizontal motions at
all scales (which are manifest in Figs.~\ref{snapshotsdiv} and \ref{urdiv}). 
At this stage, we do not have a theory for the scalings
of the energy spectrum $E_T(\ell)$ of the weaker horizontal toroidal
velocity field, which seems to be only weakly coupled to the spheroidal
divergent dynamics.

Finally, at even larger scales ($\ell<120$), beyond the
supergranulation spectral break, our observational analysis 
shows a regime change to an intriguing $E_S(\ell)\propto
\ell^2$. This might be one of the first observations of a large-scale
$k^2$ pre-injection spectrum predicted by \cite{saffman67}. In theory,
this spectrum is established as a result of long-range pressure
correlations in turbulent flows in which fluid ``eddies'' have a
random distribution of linear momentum \citep{davidson}, as opposed to
a random distribution of angular momentum, which would instead result
in a $k^4$ spectrum \citep{batchelor56}. In the solar photosphere,
such eddies could perhaps be associated with supergranulation cells
and smaller-scale convective motions. 

\section{Discussion\label{discussion}}
We have presented a set of consistent arguments suggesting 
that turbulent flows at scales larger than granulation are a
manifestation of statistically self-similar, strongly nonlinear
convection in the bulk adiabatic layer below the solar surface. 
While the classic idea that supergranulation has a convective
origin has usually been dismissed due to the seeming lack of photometric
intensity contrast at the corresponding scales at the surface
\citep{langfellner16},
helioseismology suggests that relative temperature fluctuations 
of approximately a few percent are present underneath the surface at
such scales \citep{duvall97}, in line with our calculation in 
Sect.~\ref{thestimate}. Besides, all existing large-scale simulations,
including those with radiative transfer that do not display any
``meso'' or ``super'' scale intensity contrast at
the surface, exhibit a strong buoyancy driving and flows at such scales in the
bulk of the convective layer \citep[e.~g.,][see also \cite{rieutord10a},
Fig. 12]{rincon05,nordlund09,bushby14,lord14,cossette16}. 
Importantly, in these simulations, these scales are significantly
larger than those of the most unstable linear modes of the system at
rest, and the statistical order at such scales emerges dynamically in
the nonlinear regime. If there are indeed significant thermal
fluctuations associated with supergranulation-scale convection in the
first few Mm below the surface, they may be blanketed by the thinner
thermal granulation boundary layer, or could be optically
 thick due to the extreme dependence of opacities on  temperature
 \citep{nordlund09}. Our theory also suggests that anisotropic
 supergranulation-scale convection, although it generates strong
horizontal flows, does not significantly contribute to the transport
of thermal energy due to the weakness of the
radial component of the velocity field at such scales.

We do not have a quantitative answer as to what determines the
spectral supergranulation break, and can only
offer directions. Our theory predicts that thermal
fluctuations increase with scale as $E_\theta(\ell)\propto
\ell^{-9/5}$ in the nonlinear convection regime. However, there is
a physical limit to how large such fluctuations can be, which could
determine the maximum buoyant scale at which the self-similar scaling
should break down. This limit should be related in some way to the maximum 
entropy fluctuations that can be injected into the adiabatic layer
and, therefore, to the details of granulation and of the
superadiabatic thermal boundary layer at the surface. This conclusion
appears to be in line with the recent numerical simulations of this
problem by \cite{cossette16}, which show that the thickness of the boundary
layer has a strong effect on the energy-containing scale of the 
turbulent energy spectra. However, other physical explanations 
 \citep[e.g.,][]{featherstone16} cannot be ruled out at this stage.

Based on our experience with this data, the observational 
analysis and comparison with theory presented above are
close to the limit of what is achievable with a combination of
surface tracking and direct Doppler measurements given their disparate
natures. Future progress will probably come from acoustic tomography
\citep{toomre15} and advanced numerical models. 
While somewhat speculative at this stage, our theory, including
\equs{vscaling}{tscaling}, may soon be testable using such
tools. An important question in this respect asks to what
  extent pristine power-law scalings derived from simple dynamical
  arguments can be realized in systems such as the photospheric transition
region, where a variety of physical processes become intertwined.

The qualitative implications of such analyses could be wider
than the solar context. They may notably provide us with fundamental
insight into the structure of anisotropic turbulence in general
\citep[e.g.,][]{nazarenko11}, as well as  into turbulence in more
distant, unresolved astrophysical systems supporting anisotropic waves or instabilities, 
such as stellar interiors \citep{zahn08,miesch09}, galaxy clusters
\citep{zhuravleva14} and accretion discs \citep{walker16}.

\bibliographystyle{aa}
\bibliography{sg}

\begin{acknowledgements}
This work is dedicated to the memory of Jean-Paul Zahn, 
a pioneer in astrophysical fluid dynamics research whose passion 
and work have been an invaluable source of inspiration to us.

We thank Nathana\"el Sch{\ae}ffer for his assistance with the
\texttt{SHTns} library, the SDO/HMI data provider JSOC,
the SDO/HMI team members, in particular P. Scherrer,
S. Couvidat and J. Schou for sharing information on the
calibration and removal of systematics. This work was granted access
to the HPC resources of CALMIP under allocation 2011-[P1115]. We also
thank N. Renon for his assistance with the CST code parallelization.

The work of AAS was supported in part by grants from the UK STFC and
EPSRC. FR and AAS thank the Wolfgang Pauli Institute, Vienna, for its
hospitality.

\end{acknowledgements}
\appendix

\section{Photospheric velocity field reconstruction\label{reconstruct}}

\subsection{Data reduction}
In order to obtain $(u_x,u_y,u_z)$ needed for the  reconstruction of
the three components of the photospheric velocity field in
\equs{ur}{uphi}, we used raw $4096\times4096$ HMI photometric and
Doppler images with a $0.5''$ pixel size ($\sim360\,\mathrm{km}$)
every $\delta t=45\,\mathrm{s}$. 

\subsubsection{Reorientation and derotation of images}
All the images were first reoriented using the SDO CROTA2 fits header 
keyword to have the solar north pole pointing 'up' ($y>0$
corresponds to the northern hemisphere), and derotated' and aligned
to avoid rotational smearing. We used the following rotation profile
adjusted from the raw Doppler data averaged over the 24 hours of
observations:
\begin{equation}
\label{rotlaw}
  \Omega (\lat)=2.87 - 0.41\sin^2(\lat) -
0.62\sin^4(\lat)\,\,\,\mu\mathrm{rad}\,\mathrm{s}^{-1}\!,
\end{equation}
where $\lat=\pi/2-\theta$ is the latitude.

\subsubsection{Coherent Structure Tracking (CST) analysis}
Starting with a sequence of intensity images, the CST algorithm
\citep{rieutord07,roudier12,roudier13} determines the
projection ($u_x$,$u_y$) in the CCD plane of the photospheric 
Eulerian velocity field by following the trajectories of many
individual intensity structures such as granules during their
entire ``life'' (the period of time, typically a few minutes, during
which a structure remains a coherent, single object that does not split or
merge). The average velocity of each structure is computed from its
total displacement over this time interval, and supersonic velocities
(larger than 7 $\mathrm{km}\,\mathrm{s}^{-1}$) are rejected. The
velocities obtained this way are irregularly spaced on the
field of view, and the last part of the procedure consists in
approximating the results by the best differentiable Eulerian field
fitting the data, using a multi-resolution wavelet analysis
\citep{rieutord07}.
Applying this procedure to the SDO/HMI data set, we obtained a
sequence of $586^2$ velocity-field maps with a temporal resolution of
30~min and a spatial resolution of 2.5~Mm (3.5'', or 7 HMI pixels,
corresponding to a maximum angular degree $\ell \sim 850$).
We further removed the $(x,y)$ velocity signal associated with the
satellite motion
 \begin{eqnarray}
 u_x^{\mathrm{sat}}(x,y,t) & = & -V_W(t)/D_\sun\sqrt{R_\sun^2-\left(x^2+y^2\right)}~,\\
 u_y^{\mathrm{sat}}(x,y,t) & = & -V_N(t)/D_\sun\sqrt{R_\sun^2-\left(x^2+y^2\right)}~,
 \end{eqnarray}
where $D_\sun$ is the distance from the satellite to the Sun, and
$V_W$ and $V_N$ are the westwards and northwards components of the
satellite peculiar velocity (given in the data file's headers).
We finally found it necessary to filter out all the signal at
spherical harmonics $\ell < 20$, as the latter is polluted by
systematic large-scale velocity-field residuals associated with
imperfect corrections of the rotation, satellite motions, plages
regions, and artificial large-scale drifts imprinted by the CST
analysis itself far from disc center \citep{roudier13}. This filtering
was performed in the spherical-harmonics space by means of the
harmonic-transform machinery described  App.~\ref{harmonicanalysis}.

\subsubsection{Doppler data analysis}
Dopplergrams map the out-of-plane (line-of-sight, l.o.s.) component of
the photospheric velocity field. The SDO/HMI convention is that 
$u_{\mathrm{Dop}}$ is taken as positive when the flow is away from the
observer, so that the out-of-plane velocity towards the observer is
$u_z\simeq -u_{\mathrm{Dop}}$ (Fig.~\ref{figcoord}). This equality is
only approximate because of the inclination between the
actual l.o.s  and the $\vec{e}_z$ unit vector due to the finite
distance between the Earth and the Sun. This effect is taken
into account to eliminate the global rotational satellite motion
signals, but is otherwise negligible. We first substracted the
rotational Doppler shift
\begin{equation}
u_\mathrm{Dop}^{\mathrm{rot}}=\Omega(\lambda)\,\sin\varphi\,\cos\lambda\,\cos B_0\,
R_\sun~,
\end{equation}
where $\Omega(\lambda)$ is given in \equ{rotlaw},
and the Doppler shift associated with the satellite motion
\begin{equation}
u_\mathrm{Dop}^{\mathrm{sat}}= V_R\cos\sigma-V_W\left(x/D_\sun\right)-V_N\left(y/D_\sun\right)~,
\end{equation}
where $\sigma=\arcsin\left(\sqrt{x^2+y^2}/D_\sun\right)$.
The images were subsequently 'derotated' in the same way as the
white-light images and further corrected from a polynomial radial
limbshift function adjusted from ring averages of two hours of data
taken at different heliocentric angles from disc center $\rho$,
\begin{equation}
\label{eq:limb}
 u_\mathrm{Dop}^{\mathrm{limb}} = \left(-0.35\,x+0.2\,x^2+0.46\,x^3\right)\mathrm{km}\,\mathrm{s}^{-1},
\end{equation}
where $x=1-\cos\rho$ and 
\begin{equation}
  \label{eq:rho}
  \rho = \arcsin\sqrt{\sin^2\theta\sin^2\varphi+\left(\cos\theta\cos
      B_0-\sin\theta\cos\varphi\sin B_0\right)^2}~,
\end{equation}

Since the $586^2$ velocity-field maps derived from the CST
have a 2.5~Mm spatial resolution, we then downsampled the 
the $4096^2$ Dopplergrams to $586^2$, keeping only one point out 
of seven in each direction, and averaged over 30-min periods to obtain
an effective sequence of maps of $u_z$, with the same spatial and
temporal resolution as that of the $(u_x,u_y)$ maps derived from CST.

\subsection{Noise analysis and $\ell$-$\nu$ filtering}
\label{ellnufiltering}
\begin{figure}
\includegraphics[width=\columnwidth]{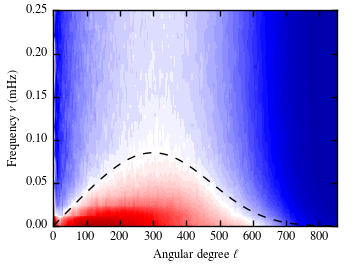}
\includegraphics[width=\columnwidth]{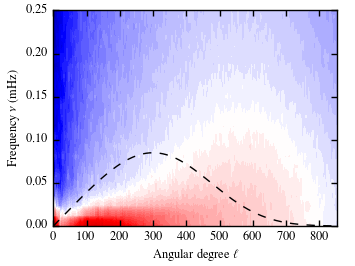}
\caption{\label{ellnudiagram}Distribution of power in the $\ell$-$\nu$
  plane for a 24-hour sequence of $u_x$ signal derived from the CST
  (top) and SDO/HMI Doppler signal (bottom). The dashed line marks the
  filtering envelope (\ref{filtenvelope}). The color-scale from blue to
  red is identical on both plots and extends over five orders of
  magnitude.}
\end{figure}

While the raw CST signal is not sensitive to solar oscillations or
granulation, it is contaminated by an undesirable intrinsic
algorithmic noise that can, to some extent, be removed
via $\ell$-$\nu$ filtering using the harmonic-transform
machinery described in App.~\ref{harmonicanalysis}.
The $\ell$-$\nu$ diagram  of the $x$-component of the raw CST
velocity field computed from 24 hours of data is shown in
Fig.~\ref{ellnudiagram}  (top -- the $y$-component behaves similarly).
The main signal (red bump)  is contaminated by a white-noise-in-time
component in the form  of a vertical white stripe. To remove this
noise, we filtered out  all the data at frequencies larger than those
given by the filtering envelope (also shown in the figure):
\begin{equation}
\label{filtenvelope}
\nu_c(\ell)=0.085\,\ell\,\mathrm{e}^{-1/3\,(\ell/300)^3}/(300\, \mathrm{e}^{-1/3})\,\,\mathrm{mHz}~.
\end{equation}

The $\ell$-$\nu$ diagram of the Doppler signal (first downgraded to
the $586\times586$ CST resolution) for the same range of frequencies
is shown in Fig.~\ref{ellnudiagram} (bottom ; the high-frequency part, not shown
here, contains the usual $p$-mode ridges). The spectral support of the
turbulent Doppler signal appears to be somewhat different from that of
the CST. In particular, there is a fairly large, relatively
high-frequency component in the $300<\ell<800$  range (in white in the
figure), which we attribute to the large-scale spectral tail of
granulation and which is not captured by the CST. Note that
the decrease of this signal at the smallest scales represented is 
due to the coarse-graining associated with the downgrade of the
Doppler maps to the CST resolution.

The presence of this extra signal component in the Doppler data but not
in the CST implies that the two raw signals cannot naively be mixed
in the 3D-velocity-field reconstruction in \equs{ur}{uphi}, as this 
would lead to spurious cross-talk between the different field
components. In our view, the most consistent way to proceed is,
therefore, to apply the same $\ell$-$\nu$ filtering envelope
(\ref{filtenvelope})  to both Doppler and CST fields. While it is
imperfect and does not totally remove the granulation tail contribution
from the large-scale signal, this procedure focuses the analysis on
large-scale motions inferred from the CST, limits the risk of
cross-analysing signals of different physical origin, and makes it
possible to extract the weak supergranulation-scale $u_r$ signal from
the Doppler data out of the  strong large-scale tail of the surface
granulation dynamics. This filtering also entirely removes the solar
oscillation signal from the Doppler data.

Finally, we found it necessary to discard the residual
signal in the spherical harmonics $\ell < 20$  because of the presence
of significant instrumental systematics in the 2010 SDO data 
\citep[private communication by Couvidat, Scherrer, Schou, see
  also][]{hathaway15}.

\section{Harmonic-analysis technique\label{harmonicanalysis}}
\subsection{Harmonic transforms and spectra}
Any sufficiently smooth scalar field $\psi(\theta,\varphi)$ on the sphere
can be expanded in terms of spherical harmonics as
\begin{equation}
\psi(\theta,\varphi)=\sum_\ell\sum_{m=-\ell}^\ell \psilm\, \ylm (\theta,\varphi)~,
\end{equation}
where $\ylm$ is the orthonormalized spherical
harmonic of degree $\ell$ and azimuthal wavenumber $m$, and
\begin{equation}
  \label{eq:psilm}
  \psilm=\displaystyle{\int_0^{2\pi}\int_0^\pi \psi(\theta,\varphi){\ylm}^*(\theta,\varphi)\sin\theta\, d\theta\, d\varphi}~.
\end{equation}
The one-dimensional energy spectrum of $\psi$ is defined as
\begin{equation}
\label{eq:psilmspectrum}
E_\psi(\ell)=\sum_{m=-\ell}^\ell |\psilm|^2.
\end{equation}
This decomposition can be generalized to vector fields, as explained
in the main article; we refer to \equ{vectorexpansion}. The rest of this Appendix
describes the procedure used to compute  the harmonic transforms of
scalar and vector fields derived from  SDO/HMI data sets. 

\subsection{Data interpolation on a Gauss-Legendre-Fourier grid}
Spherical-harmonics transforms of all fields are computed using
Gauss-Legendre quadrature coupled to a Fast-Fourier-Transform
algorithm in the azimuthal direction. This requires the knowledge of
the field on a two-dimensional (2D) Gauss-Legendre-Fourier (GLF) grid of size
$N_\theta\times N_\varphi$ of polar angles corresponding
to Gauss nodes $\theta_n$ $(0<\theta_n<\pi, 1\leq n \leq N_\theta)$, 
and equally-spaced longitudes $\varphi_p=2\pi p/N_\varphi$ (with
$0\leq p < N_\varphi$; by convention, $\varphi=0$ is the central
meridian of the visible disc, $\varphi=\pi/2$ is the eastern limb
and  $\varphi=-\pi/2$ is the western limb). However, our fields
are originally available on a $586^2$ Cartesian 2D grid. Therefore,
interpolation from the original Cartesian grid to a GLF grid is a
prerequisite. To carry it out, we choose a spectral resolution
$(N_\theta,N_\varphi)$, compute the Cartesian coordinates of the
corresponding GLF grid in the plane of the
sky/CCD, and interpolate the field on the original,
regular Cartesian grid to these coordinates using the
\texttt{RectBivariateSpline} function of the Python module
\texttt{scipy.interpolate}. A similar, reverse interpolation procedure
is used to reproject fields manipulated in spectral space from the GLF
grid to the original ``observation'' grid.  This is, for
instance, required to obtain the divergence field on this grid.

\subsection{Estimator of spectral coefficients}
An apodizing window $W(\theta,\varphi)$ must be applied to the data in
order to deal with the fact that we only see one side of the Sun, and
to eliminate near-limb regions where the CST analysis is unreliable
due to projection effects. We use the estimator 
 \begin{equation}
\label{eq:spectralcoeff}
\psilmtilde=\f{1}{\Nlm}\displaystyle{\int_0^{2\pi}\int_0^\pi
    \psi(\theta,\varphi)W(\theta,\varphi){\ylm}^*(\theta,\varphi)\sin\theta\,
    d\theta\, d\varphi}
\end{equation}
to obtain  the spectral coefficients $\psilm$ of a scalar field $\psi$
known on a limited area (see \cite{hathaway92} for a similar
definition ; the harmonic coefficients of vector fields can be
estimated similarly). The normalization coefficients $\Nlm$ are
functionally dependent on the apodizing window, and can be defined in
different ways (see next paragraph). The integral in the numerator of
\equ{eq:spectralcoeff}, or its generalization to the vector case, are
computed from the discrete data sets interpolated on the GLF grid using 
the python interface of the \texttt{SHTns} library
\citep{schaeffer13}. The tilde notation for the estimators of the
spectral coefficients is dropped in the text to simplify notations.

\subsection{Apodizing and normalization issues\label{appwindow}}
Dealing with data on a limited area, or apodizing it, introduces
leakage between different harmonics.
In helioseismology, this leakage can partly be alleviated by
separating the oscillatory components of the signal according to their 
discrete temporal frequencies \citep{schou94,howe98}. However, this
approach cannot be used here because the flows that we analyze are
characterized by  a continuum of time scales. As a result, leakage
cannot be eliminated and can notably lead to overestimating
the magnitude of the flow. 

The results of a detailed technical analysis including careful 
tests and validations using synthetic data (not presented here 
but available on request) show that a sensible approach to
estimate accurately the spectral coefficients in our problem 
is to use a ``$\rho$ (heliocentric angle from disc center) window'' function,
\begin{equation}
  \label{eq:rhomask}
W(\rho)=\left[\f{1}{2}\left(1-\tanh S\left(\f{\rho-\rho_c}{\rho_c}\right)\right)\right]^n~,  
\end{equation}
where $S$ is a smoothing parameter,  $\rho_c$ is the critical
heliocentric angle from disc center above which the apodization
becomes significant, and integer $n$ parametrizes the sharpness of the
fall-off of the window function. We adopt an ``energy-conserving'' 
normalization, which consists in directly calibrating the coefficients
$\Nlm[W]$ introduced in \equ{eq:spectralcoeff}
once and for all based on a comparison between the
(known) analytical spectra of a few test synthetic scalar fields
defined over the whole sphere on the one hand, and the spectra
obtained through a (unrenormalized) harmonic analysis of the apodized
versions of the same fields on the other hand. For the family of $W$
given by \equ{eq:rhomask}, we found that using a normalization
factor independent of $\ell$ and $m$ is appropriate. Most importantly,
although this strategy does not entirely eliminate the leakage
problem, it does alleviate that of overestimating the spectral amplitude.

\subsection{Harmonic-analysis parameters}
All the results of the paper involving a harmonic analysis were
computed for a resolution $(N_\theta,N_\varphi)=(846,1728)$,
corresponding to $\ell_\mathrm{max}=m_\mathrm{max}=845$. This
resolution was determined using the Nyquist criterion applied to the
$586^2$ CST grid resolution of $\sim 2.5$~Mm. We used the window function
\begin{equation}
  \label{eq:window60}
W_{60}(\rho)=\left[\f{1}{2}\left(1-\tanh 5\left(\f{\rho-\pi/3}{\pi/3}\right)\right)\right]^6~,
\end{equation}
which apodizes the field sharply at angular distances from disc 
center larger than $\rho_c=60^{\degree}$, and the same window with a
narrower opening $\rho_c=15^{\degree}$. The normalization of the harmonic
spectra of the apodized data is based  on the energy-conserving
normalization introduced above, with $\Nlm[W_{60}]=0.36$ for all
$\ell$ and $m$.

 \end{document}